\documentstyle[twoside,12pt]{article}

\title{Bilinear Structure and Exact Solutions of
       the Discrete Painlev\'e I Equation}

\author{
\vspace{10pt}\\
 {Yasuhiro Ohta}\\
 {\it Department of Applied Mathematics, Faculty of Engineering,}\\
 {\it Hiroshima University,}\\
 {\it 1-4-1 Kagamiyama, Higashi-Hiroshima, Hiroshima 724, Japan}\\
 {ohta@kurims.kyoto-u.ac.jp}\\
\vspace{3pt}\\
 {Kenji Kajiwara}\\
 {\it Department of Electrical Engineering, Doshisha University,}\\
 {\it Tanabe, Kyoto 610-03, Japan}\\
 {kkajiwar@duaic.doshisha.ac.jp}\\
\vspace{3pt}\\
 {and}\\
\vspace{3pt}\\
 {Junkichi Satsuma}\\
 {\it Department of Mathematical Sciences, University of Tokyo,}\\
 {\it 3-8-1 Komaba, Meguro-ku, Tokyo 153, Japan}\\
 {satsuma@tansei.cc.u-tokyo.ac.jp}
\vspace{10pt}\\}

\date{}

\begin{document}

\maketitle

\pagestyle{myheadings}
\markboth{Y. OHTA, K. KAJIWARA AND J. SATSUMA}
         {EXACT SOLUTIONS OF
	  DISCRETE PAINLEV\'E I EQUATION}

\footnotetext[1]{35Q58, 58F07}

Nonlinear integrable discrete systems
have recently attracted much attention.
Among them,
the most fundamental ones are
the discrete analogues of Painlev\'e equations.
Some properties such as the existence of Lax pair,
B\"acklund transformation and singularity confinement property,
have been studied for these discrete Painlev\'e equations\cite{GR}.
However as for the solutions,
only in a few cases particular solutions have been derived
and the structure of the solution space is still almost unknown.
We have studied the solution in determinant form
for the discrete Painlev\'e II equation
based on the bilinear formalism\cite{KOSGR}.
We here investigate the bilinear structure for
the discrete Painlev\'e I (dPI) equation on semi-infinite lattice,
and show that its solutions are given in terms of
the Casorati determinant.

\ \par
The dPI equation on semi-infinite lattice,
$$
u_{N+1} + u_N + u_{N-1} = a + {bN \over 2u_N},
\quad
\hbox{for }N\ge1,
\qquad
u_0 = 0,
\eqno(1)
$$
where $a$ and $b$ are arbitrary constants,
appears in the theory of two-dimensional quantum gravity\cite{DS}.
The partition function of the matrix model,
$$
Z = \int {\rm d}M \exp\bigl(-{\rm tr}V(M)\bigr),
\eqno(2)
$$
$$
V(M) = c_1 M^2 + c_2 M^4,
\eqno(3)
$$
where the integral is taken over all $m\times m$ Hermite matrix $M$
with the Haar measure,
is given in terms of the solution of dPI eq.(1).
Actually we have
$$
Z = m! w^m {u_1}^{m-1}{u_2}^{m-2}\cdots u_{m-1},
\qquad
w = \int_{-\infty}^\infty {\rm d}x \exp\bigl(-V(x)\bigr),
$$
where $u_N$ is the solution of (1) with the boundary condition,
$$
u_1 = {1 \over w}\int_{-\infty}^\infty {\rm d}x x^2\exp\bigl(-V(x)\bigr),
$$
and $a$, $b$ and $c_1$, $c_2$ are related as
$$
a = - {c_1 \over 2c_2},
\qquad
b = {1 \over 2c_2}.
$$
We expect that investigation of solutions
of discrete Painlev\'e equations is useful to understand
the structure of physical systems
such as the above matrix model.

\ \par
In the context of the bilinear formalism,
the dPI eq.(1) is naturally extended to the discrete coupled system,
$$
\cases{
v_N^n + u_N^n + v_{N-1}^n
 = a + {\displaystyle bN \over \displaystyle u_N^n},
 &for $N\ge1$,\cr
\noalign{\vskip3pt}\cr
u_{N+1}^n + v_N^n + u_N^n
 = a + {\displaystyle b(n+N)+c \over \displaystyle v_N^n},
 &for $N\ge0$,}
\eqno(4)
$$
with the boundary condition,
$$
u_0^n = 0.
$$
In (4),
$n$ is the auxiliary suffix introduced to have a natural expression
for the solutions in bilinear formalism.
The coupled system (4) reduces to the dPI eq.(1)
by taking $c=b/2$, $n=0$ and
$$
u_{2N} = u_N^0,
\quad
u_{2N+1} = v_N^0,
\quad
\hbox{for }N\ge0.
$$
In the following,
we investigate the solution of this extended dPI equations
using the bilinear method.
The bilinear method is originally developed
to solve the soliton equations\cite{H}.
Recently it has been shown that the method is also
applicable to the discrete Painlev\'e II equation\cite{KOSGR}.

First we introduce the $\tau$ function
by the dependent variable transformation,
$$
u_N^n = {\tau_{N+1}^n\tau_{N-1}^{n+1} \over \tau_N^{n+1}\tau_N^n},
\qquad
v_N^n = {\tau_{N+1}^{n+1}\tau_N^n \over \tau_{N+1}^n\tau_N^{n+1}}.
$$
Substituting these into (4) and decoupling the resultant equations,
we arrive at the bilinear form for the extended dPI equations,
$$
\tau_{N+1}^{n-1}\tau_{N-1}^{n+1}
 = \tau_N^{n-1}\tau_N^{n+1} - \tau_N^n\tau_N^n,
\eqno(5)
$$
$$
\tau_{N+1}^n\tau_{N-1}^n
 = \bigl(b(n+N-1)+c\bigr)\tau_N^n\tau_N^n
 - \bigl(b(n-1)+c\bigr)\tau_N^{n+1}\tau_N^{n-1},
\eqno(6)
$$
$$
\tau_{N+1}^{n+1}\tau_N^{n-1} - \tau_{N+2}^{n-1}\tau_{N-1}^{n+1}
 = a\tau_{N+1}^n\tau_N^n
 + \bigl(b(n-1)+c\bigr)\tau_{N+1}^{n-1}\tau_N^{n+1}.
\eqno(7)
$$
Equations (4) actually follows from (5)-(7).
For example we derive the first equation in (4).
Dividing (7) by $\tau_{N+1}^n\tau_N^n$,
we get
$$
{\tau_{N+1}^{n+1}\tau_N^{n-1} \over \tau_{N+1}^n\tau_N^n}
 - {\tau_{N+2}^{n-1}\tau_{N-1}^{n+1} \over \tau_{N+1}^n\tau_N^n}
 = a
 + \bigl(b(n-1)+c\bigr)
   {\tau_{N+1}^{n-1}\tau_N^{n+1} \over \tau_{N+1}^n\tau_N^n},
$$
which is rewritten by using (5) as
$$
v_N^n + u_N^n
 = a
 + \bigl(b(n-1)+c\bigr)
   {\tau_{N+1}^{n-1}\tau_N^{n+1} \over \tau_{N+1}^n\tau_N^n}.
\eqno(8)
$$
Eliminating the term $\tau_N^{n+1}\tau_N^{n-1}$ from (5) and (6),
we have
$$
\tau_{N+1}^n\tau_{N-1}^n
 = bN\tau_N^n\tau_N^n
 - \bigl(b(n-1)+c\bigr)\tau_{N+1}^{n-1}\tau_{N-1}^{n+1}.
$$
Multiplying this by
$\tau_N^{n+1}/(\tau_{N+1}^n\tau_N^n\tau_{N-1}^{n+1})$,
we get
$$
v_{N-1}^n
 = {bN \over u_N^n}
 - \bigl(b(n-1)+c\bigr)
   {\tau_{N+1}^{n-1}\tau_N^{n+1} \over \tau_{N+1}^n\tau_N^n}.
\eqno(9)
$$
Adding (8) and (9), we obtain the first equation in (4).
The second equation in (4) is similarly derived from (5)-(7).
Finally the boundary condition $u_0^n=0$ is satisfied
by taking $\tau_{-1}^n=0$.

Now we solve the bilinear difference equations (5)-(7).
The solution is given as
$$
\tau_N^n = \left|\matrix{
 A_n       &A_{n+1} &\cdots &A_{n+N-1} \cr
 A_{n+1}   &A_{n+2} &\cdots &A_{n+N}   \cr
 \vdots    &\vdots  &       &\vdots    \cr
 A_{n+N-1} &A_{n+N} &\cdots &A_{n+2N-2}}\right|,
\quad
N\ge1,
\eqno(10)
$$
$$
\tau_0^n = 1,
\qquad
\tau_{-1}^n = 0,
$$
where $A_n$ is a discrete Airy function satisfying
$$
A_{n+2} = aA_{n+1} + (bn+c)A_n.
\eqno(11)
$$
It is proved by using the standard technique\cite{KOSGR}
that the above $\tau_N^n$ actually satisfies (5)-(7).
As an example, we give the proof for (6).
In the Casorati determinant (10),
adding $(N-1)$-th column multiplied by $-a$ to $N$-th column,
adding $(N-2)$-th column multiplied by $-b(N-2)$ to $N$-th column
and using (11),
we get
$$
\tau_N^n = \left|\matrix{
 A_n       &A_{n+1} &\cdots &A_{n+N-2}  &\bigl(b(n-1)  +c\bigr)A_{n+N-3} \cr
 A_{n+1}   &A_{n+2} &\cdots &A_{n+N-1}  &\bigl(bn      +c\bigr)A_{n+N-2} \cr
 \vdots    &\vdots  &       &\vdots     &\vdots                          \cr
 A_{n+N-1} &A_{n+N} &\cdots &A_{n+2N-3} &\bigl(b(n+N-2)+c\bigr)A_{n+2N-4}}
\right|.
$$
Furthermore,
adding $(N-j-1)$-th column multiplied by $-a$ to $(N-j)$-th column
and adding $(N-j-2)$-th column multiplied by $-b(N-j-2)$
to $(N-j)$-th column,
for $j=1,\cdots,N-2$ successively,
we have
\begin{eqnarray*}
&{}&{\tau_N^n \over
  \bigl(b(n-1)+c\bigr)\bigl(bn+c\bigr)\cdots\bigl(b(n+N-2)+c\bigr)} \\
&{}&= \left|\matrix{
 {1 \over b(n-1)  +c}A_n       &A_{n-1}   &\cdots &A_{n+N-3} \cr
 {1 \over bn      +c}A_{n+1}   &A_n       &\cdots &A_{n+N-2} \cr
 \vdots                        &\vdots    &       &\vdots    \cr
 {1 \over b(n+N-2)+c}A_{n+N-1} &A_{n+N-2} &\cdots &A_{n+2N-4}}\right|.
\end{eqnarray*}
Applying the Jacobi formula to the above determinant,
we obtain the bilinear equation,
\begin{eqnarray*}
&{}&{\tau_N^n \over
  \bigl(b(n-1)+c\bigr)\bigl(bn+c\bigr)\cdots\bigl(b(n+N-2)+c\bigr)}
 \tau_{N-2}^n \\
&{}&= {\tau_{N-1}^n \over
    \bigl(b(n-1)+c\bigr)\bigl(bn+c\bigr)\cdots\bigl(b(n+N-3)+c\bigr)}
   \tau_{N-1}^n \\
&{}&- {\tau_{N-1}^{n+1} \over
    \bigl(bn+c\bigr)\bigl(b(n+1)+c\bigr)\cdots\bigl(b(n+N-2)+c\bigr)}
   \tau_{N-1}^{n-1},
\end{eqnarray*}
which is essentially the same as (6).
Hence we have shown
that $\tau_N^n$ given by (10) satisfies the bilinear difference equation (6).
The other equations are also proved in a similar way.

We note that the discrete Airy function
defined by (11) is expressed as
$$
A_n = (-b)^{(n-1+c/b)/2}
      H_{n-1+c/b}\left({a \over (-b)^{1/2}}\right),
$$
where $H_\nu(x)$ is the Hermite-Weber function
satisfying the contiguity relation,
$$
H_{\nu+1}(x) - xH_\nu(x) + \nu H_{\nu-1}(x) = 0.
$$

\ \par
We have investigated the Casorati determinant solution
of the extended dPI equations.
As was mentioned,
the solution for dPI equation gives
the partition function of a matrix model.
Hence a natural question is
what kind of matrix model corresponds to the extended dPI eqs. (4).
We find that it is the matrix model with the potential,
$$
V(M) = c_1 M^2 + c_2 M^4 + c_0 \log M^2.
\eqno(12)
$$
Actually the partition function defined in (2) with the potential (12)
is given as
$$
Z = m! w^m (v_0^0)^{m-1}(u_1^0)^{m-2}(v_1^0)^{m-3}(u_2^0)^{m-4}\cdots,
$$
where $u_N^0$ and $v_N^0$ are the solution of (4)
with the boundary condition,
$$
v_0^0 = {1 \over w}\int_{-\infty}^\infty {\rm d}x x^2\exp\bigl(-V(x)\bigr),
$$
and the parameters,
$$
a = - {c_1 \over 2c_2},
\qquad
b = {1 \over 2c_2},
\qquad
c = {1 - 2c_0 \over 4c_2}.
$$

Finally we comment on the continuous limit of dPI.
The dPI equation itself reduces to the Painlev\'e I equation
on some condition for the parameters.
However, the relation between the solution given in this article
and that for the Painlev\'e I equation is not clear yet,
because the $\tau$ function (10) is considered
only on the semi-infinite lattice.

\vspace{20pt}


\end{document}